\documentclass[%
 preprint, 
 amsmath,amssymb,
 aps, physrev,
]{revtex4-2}

\usepackage{newtxtext,newtxmath}
\usepackage{amsmath}
\usepackage{lineno}
\usepackage{ulem}
\usepackage{color}
\usepackage{soul, color, xcolor}

\usepackage{graphicx}
\usepackage{dcolumn}
\usepackage{bm}

\begin{document}

\title{\textbf{Temporal Beam Self-Cleaning in Second-Harmonic Generation} 
}%

\author{Siyu Chen,$^{1}$ Jun Ye,$^{1,2,3,*}$ Lei Du,$^{1}$ Wenwen Cheng,$^{1}$ Jiangming Xu,$^{1}$ Rongtao Su,$^{1,2,3,*}$ Pu Zhou$^{1,}$}%
\email{Contact author: yejun1021@163.com, surongtao@126.com, zhoupu203@163.com}
\author{Zongfu Jiang$^{1,2,3}$}
\affiliation{%
 $^{1}$College of Advanced Interdisciplinary Studies, National University of Defense Technology, Changsha 410073, China\\
 $^{2}$Nanhu Laser Laboratory, National University of Defense Technology, Changsha 410073, China\\
 $^{3}$Hunan Provincial Key Laboratory of High Energy Laser Technology, National University of Defense Technology, Changsha 410073, China
}%

\begin{abstract}
The spatio-temporal beam quality of laser sources is crucial for applications such as nonlinear spectroscopy and master oscillator power amplification (MOPA) systems. However, the temporal stability remains challenged by issues like linewidth broadening and high-power demand in efforts to improve it. In this work, we investigate the effect of the second-harmonic generation (SHG) process on the laser characteristics under three longitudinal mode regimes: single-longitudinal-mode (SLM), dual-longitudinal-mode (DLM), and multi-longitudinal-mode (MLM). The results demonstrate that the SHG process effectively stabilizes the temporal characteristics of the laser and enhances its correlation, leading to a temporally clean output beam. The physical mechanism of the observed temporal stabilization effect can be attributed to a high-peak-pulse attenuation effect, jointly induced by nonuniform longitudinal-mode depletion and phase preservation in the residual fundamental wave. Statistical analysis indicates that at the maximum fundamental wave (FW) power in the MLM regime, the standard deviation and peak-to-valley values derived from the normalized temporal profile decrease from 0.6122 and 5.6846 for the FW to 0.189 and 0.8847 for the residual FW. Meanwhile, the background level of the intensity autocorrelation function (ACF) rises from $\thicksim$0.72 to $\thicksim$0.96, revealing its evolution toward a more coherent state. To the best of our knowledge, this research presents the first demonstration of laser temporal stabilization and correlation enhancement via SHG. It not only deepens the comprehension of SHG mechanisms, but also opens up a new avenue for realizing temporal beam self-cleaning of light.
\end{abstract}

 \maketitle


\section{\label{sec:level1}Introduction}

Enhancing the spatio-temporal beam quality of a light source is essential for achieving high performance in applications including nonlinear spectroscopy, MOPA systems and optical frequency combs \cite{cao2023, stihler2020, jiang2020}. Spatially, the Kerr-effect-induced beam self-cleaning phenomenon allows optical energy in multimode fibers to spontaneously transfer to the fundamental mode, resulting in a stable fundamental-mode output \cite{krupa2017, tegin2020, zhang2021}. Motivated by this observation, researchers are exploring whether an analogous “self-cleaning” mechanism can occur in the temporal domain, aiming to mitigate the detrimental impact of temporal instabilities on system performance. For instance, in MOPA systems, the temporal instability of the oscillator significantly affects the stability and performance of the system \cite{wang2020a}. Specifically, temporal instability in the seed laser exacerbates nonlinear effects during amplification \cite{raoModellingCharacterizationTime2024}, lowers the threshold for transverse mode instability (TMI) \cite{stihler2020}, and ultimately deteriorates the overall system performance. Currently, temporally stable laser sources can be realized using free-running cavity lasers without resonant cavity structures, such as superfluorescent fiber source \cite{yeSpectralBroadeningRecompression2021,yeSpectrumManipulableHundredWattLevelHighPower2019,heHighPowerSpectrumtailorable2024} and random fiber laser \cite{turitsynRandomDistributedFeedback2010a,churkinRecentAdvancesFundamentals2015,hanSpectralManipulationsRandom2024}. The structural characteristic of lacking a well-defined resonant cavity enables these sources essentially free of distinct longitudinal modes \cite{han2025, vatnik2019}, effectively suppress relaxation oscillations and self-pulsing phenomena \cite{duTemporallyStableRandom2015}, and ultimately yield a temporally stable laser output. In addition to completely eliminate the resonant cavity, another approach involves constructing a weak-feedback resonator—employing an ultralow-reflectivity fiber Bragg grating as the output cavity mirror—to effectively regulate longitudinal modes, realize modeless laser radiation, and thereby achieve a temporally stable laser output \cite{li2023a}. Furthermore, external feedback introduced outside the cavity has been employed to improve the temporal characteristics of laser source. By introducing a low-reflectivity fiber Bragg grating externally as feedback, temporal intensity fluctuations can be significantly suppressed \cite{xuEliminationSelfmodelockingPulses2017,chenPowerScalingRaman2020}. Moreover, researchers enhanced the temporal stability of laser sources by constructing a composite cavity to provide optical feedback \cite{liuUnifiedModelSpectral2021,zhangSuppressingStimulatedRaman2021}. By connecting an optical coupler externally, with one port connected to a long piece of passive fiber for external feedback and the other port for laser output, the temporal stability of the laser was effectively improved \cite{raoModellingCharacterizationTime2024,tianNovelStructureRaman2023}. Building on this approach, researchers implemented direct insertion of a passive fiber between the seed laser and amplifier to enhance temporal characteristics for subsequent amplification in the MOPA system \cite{liu32KWSingle2024,liSuppressionStimulatedRaman2019}. Nonetheless, the improvement in temporal stability achieved by these approaches comes at the expense of substantial laser linewidth broadening \cite{tianNovelStructureRaman2023}, which significantly restricts their applications in fields such as coherent beam combining and molecular physics \cite{liuDevelopment2019,corato-zanarellaWidely2023}.

Recently, the manipulation of the temporal characteristics of laser sources through nonlinear effects has emerged as a prominent research focus\cite{wenOriginSBS2025,liuEmergingLow2017,turitsynaLaminar2013}. Reverse saturable absorption (RSA) \cite{choObservation2017,yangFluorinated2005}, as a typical nonlinear phenomenon, is characterized by an increase in absorption coefficient as optical intensity rises. Such intensity-dependent property offers a novel approach for suppressing temporal fluctuations and enhancing temporal stability of the laser sources. Nevertheless, RSA implementations relying on specific materials (e.g., metalloporphyrins \cite{senge2007}, Pt(II) acetylides \cite{zhou2007,zhou2009}, graphenes \cite{lim2011,xu2009}) often encounter inherent limitations such as high activation thresholds and complex energy level structures \cite{hirata2014,zhu2023}, which hinder their practical applications. To overcome these challenges, this work proposes a novel strategy for temporal stability enhancement: inspired by the concept of achieving ``spatial self-cleaning" via the Kerr effect, we utilize the SHG process—a representative second-order nonlinear effect—to realize ``temporal self-cleaning" of laser sources. This scheme provides a new pathway for designing laser sources with high temporal stability. Furthermore, the SHG process drives the laser source towards a more coherent state, enabling simultaneous optimization of temporal stability and correlation properties.

In this work, we for the first time demonstrate that the SHG process can be harnessed to achieve temporal beam self-cleaning of light—manifested as improved temporal stability of a laser source. Leveraging a gain-related mode selection mechanism in a fiber laser, we generated laser outputs operating in the SLM, DLM, and MLM regimes. The generated outputs were subsequently directed into a nonlinear crystal to investigate the influence of the SHG process on the temporal and statistical characteristics of lasers operating in the three longitudinal mode regimes. Comparative analysis of the temporal profiles between the fundamental wave (hereafter referred to as FW) and the residual fundamental wave (hereafter referred to as residual FW in the text and Res. FW in figure legends) reveals that the SHG process significantly stabilizes the temporal characteristics of the laser source across three longitudinal mode regimes, with the stabilization effect being most substantial in the MLM regime. This conclusion is substantiated by the observation that the residual FW exhibits reduced standard deviation and peak-to-valley values of normalized temporal profiles relative to the FW. The physical mechanism behind this temporal stability effect can be attributed to the combined action of the different attenuation of the longitudinal modes and their phase preservation in the SHG process. Statistical analysis based on the probability density function (PDF) and ACF further confirms the enhanced temporal stability of the residual FW and indicates its evolution towards coherent state. This study could deepen the understanding of the SHG process and proposes a novel approach for utilizing the SHG process to realize optimization of laser temporal stability and correlation properties.

\section{\label{sec:level1}Experimental setup}
\begin{figure}[b]
\includegraphics{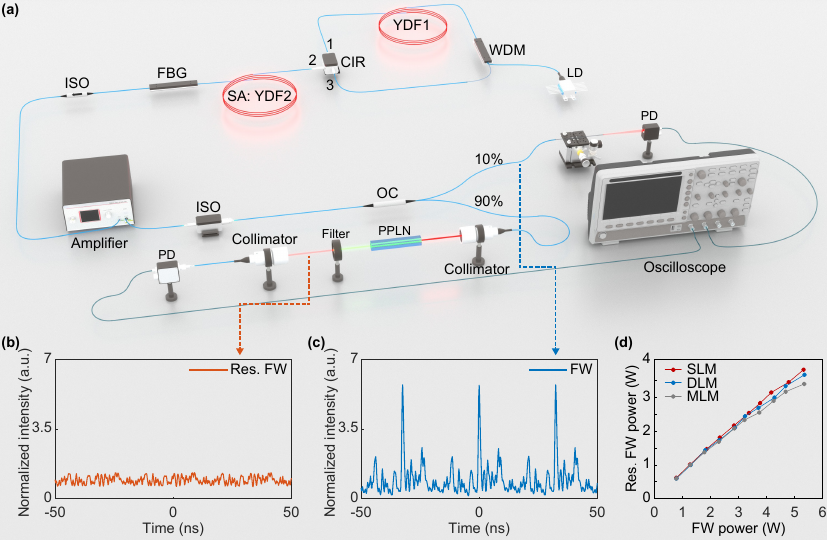}
\caption{\label{fig1} (a) Experimental setup for the impact of SHG process on laser’s characteristics. LD, laser diode; WDM, wavelength division multiplexer; YDF, Yb-doped fiber; CIR, circulator; SA, saturable absorber; FBG, fiber Bragg grating; ISO, isolator; OC, optical coupler; PPLN, periodically poled lithium niobate; PD, photodetector. Schematic diagrams of the normalized temporal profiles for the (b) residual FW and the (c) FW. (d) Variation of residual FW power with FW power.}
\end{figure}

The experimental setup for investigating the impact of the SHG process on the characteristics of the laser source is shown in Fig.~\ref{fig1}(a), comprising a fiber laser (including a seed laser and an amplifier) and a SHG unit. By leveraging the gain-related mode selection in the seed laser, we achieved transitions between SLM, DLM, and MLM regimes, the underlying principle of which is detailed in our previous work \cite{chen2026}. The seed laser is a self-built, polarization-maintaining fiber ring laser operating at 1064 nm. A 1 m-long ytterbium-doped fiber (YDF1) served as the gain fiber, and an optical circulator (CIR) is incorporated thereafter to enforce unidirectional lasing operation. An additional 1 m-long Yb-doped fiber (YDF2) functioned as a saturable absorber to facilitate the longitudinal mode selection mechanism. Specifically, the interference between the forward-propagating laser from port 2 of the CIR and the backward-propagating laser reflected by a fiber Bragg grating (FBG centered at 1064 nm with a reflectivity of 90$\%$) established a standing wave within YDF2, thereby inducing a dynamic grating. The FBG extracts 10$\%$ of the intracavity power ($\sim$1 mW) as the seed laser, which is subsequently amplified to approximately 6 W by a two-stage amplification system. The amplified laser is split by a 90:10 optical coupler, with 10$\%$ of the laser directed for characterizing the amplified fundamental wave, while the remaining 90$\%$ is used for subsequent frequency doubling. The 90$\%$ fundamental wave undergoes the SHG process in a 25-mm-long periodically poled lithium niobate (PPLN) crystal with a nonlinear coefficient of 11.8 pm/V. Following the PPLN crystal, a filter is used to eliminate the second-harmonic (SH) light, ensuring the residual light consists solely of the unconverted residual FW. The residual FW is then coupled into a single-mode fiber via a reverse-operated collimator (used as a coupler) for measurement. The temporal profiles of the FW and residual FW are simultaneously captured by two photodiodes (PDs) with a bandwidth of 5 GHz and displayed in real-time on a digital oscilloscope. Figures~\ref{fig1}(b) and \ref{fig1}(c) present schematic diagrams of the normalized temporal profiles for the FW and the residual FW, respectively, demonstrating a significant improvement in the temporal characteristics of the residual FW after the SHG process. Figure~\ref{fig1}(d) shows residual FW power versus FW power under three longitudinal mode regimes. At higher FW powers, the MLM regime exhibits the lowest residual FW power, indicating maximum FW consumption and SHG efficiency among the three regimes.

\section{\label{sec:level1}Results and discussion}
\begin{figure}[b]
\includegraphics{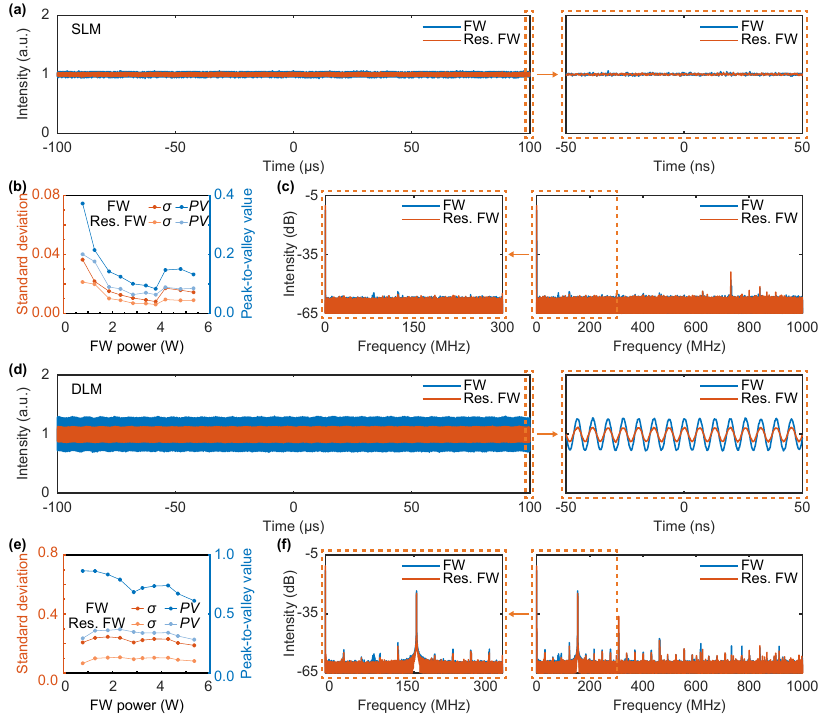}
\caption{\label{fig2} Characteristics of FW and Res. FW for (a-c) SLM and (d-f) DLM regimes. Normalized temporal profiles of the FW and Res. FW in the range of 200 $\mu$s (left) and 100 ns (right) for (a) SLM and (d) DLM regimes. Standard deviations ($\sigma$) and peak-to-valley (\(PV\)) values of the FW and Res. FW varies with FW powers for (b) SLM and (e) DLM regimes. RF spectra of the FW and Res. FW for (c) SLM and (f) DLM regimes. The inset of (c) and (f) shows the detailed RF spectra spanning the 0-300 MHz band.}
\end{figure}

The influence of SHG on the characteristics of the fundamental wave is investigated. By varying the pump current of the seed laser, fundamental wave output in three longitudinal mode regimes (SLM, DLM, and MLM) is achieved, corresponding to pump currents of 240 mA, 270 mA, and 300 mA, respectively. By measuring the characteristics of both the FW and the residual FW across these three regimes, the effects of SHG on fundamental wave characteristics are investigated. When the pump current is set to 240 mA, the laser operated in the SLM regime. Figures~\ref{fig2}(a)-\ref{fig2}(c) illustrate the spectral-temporal characteristics and statistical properties of the FW and the residual FW. Figure~\ref{fig2}(a) displays the normalized temporal profiles of the FW and the residual FW at the maximum FW power of 5.32 W, indicating a slight improvement in the temporal stability of the residual FW compared to the FW. Two statistical parameters are employed to characterize the temporal stability of laser: the standard deviation (\(Std\)) and the peak-to-valley value (\(PV\), defined as \(I_{max} - I_{min}\)) calculated from normalized temporal profiles. Here, \(I_{max}\) and \(I_{min}\) represent the maximum and minimum values of the normalized temporal intensity, respectively. Lower values of the \(Std\) and \(PV\) indicate smaller fluctuations and superior stability of the temporal intensity. The \(Std\) and \(PV\) values at different FW powers are plotted in Fig.~\ref{fig2}(b), demonstrating that SHG-processed residual FW exhibits reduced temporal fluctuations, confirming its stabilizing effect on temporal characteristics. Specifically, at the maximum FW power of 5.32 W, the \(Std\) decreased from 0.0143 for the FW to 0.0089 for the residual FW, and the \(PV\) value decreased from 0.132 to 0.0848. By performing Fourier transform on the temporal domain of the FW and residual FW, we obtained the corresponding radio frequency (RF) spectrum as shown in Fig.~\ref{fig2}(c). The absence of beat signals in the RF spectra of both the FW and residual FW indicates that the laser maintains SLM regime throughout the SHG process.

When the pump current is increased to 270 mA, the laser operated in the DLM regime, and the results showing the influence of SHG on the fundamental wave under DLM regime are presented in Figs.~\ref{fig2}(d)-\ref{fig2}(f). Figure~\ref{fig2}(d) displays the normalized temporal profiles of both the FW and the residual FW at 5.35 W FW power, clearly demonstrating the temporal stabilization effect of SHG on the DLM fundamental wave. Focusing on the 100-ns normalized temporal profile in the right panel of Fig.~\ref{fig2}(d), it reveals that increased FW intensities lead to stronger absorption during the SHG process, which consequently enhances the temporal stability of the output. The \(Std\) and \(PV\) values of normalized temporal profiles under different FW power in Fig.~\ref{fig2}(e) reveal reduced fluctuations in residual FW. At the peak FW power of 5.35 W, the SHG process reduced the \(Std\) and \(PV\) value from 0.1875 and 0.612 for the FW to 0.0814 and 0.2831 for the residual FW, respectively. The data in Figs.~\ref{fig2}(d) and \ref{fig2}(e) indicate that SHG process significantly stabilizes temporal characteristics of DLM fundamental wave, providing evidence for the realization of temporal beam self-cleaning via SHG. Figure~\ref{fig2}(f) shows the RF spectra at maximum FW power for DLM regime. The residual FW exhibits slightly reduction in beat frequency amplitude relative to the FW, suggesting partial temporal stabilization achieved via the SHG process.

\begin{figure}[b]
\includegraphics{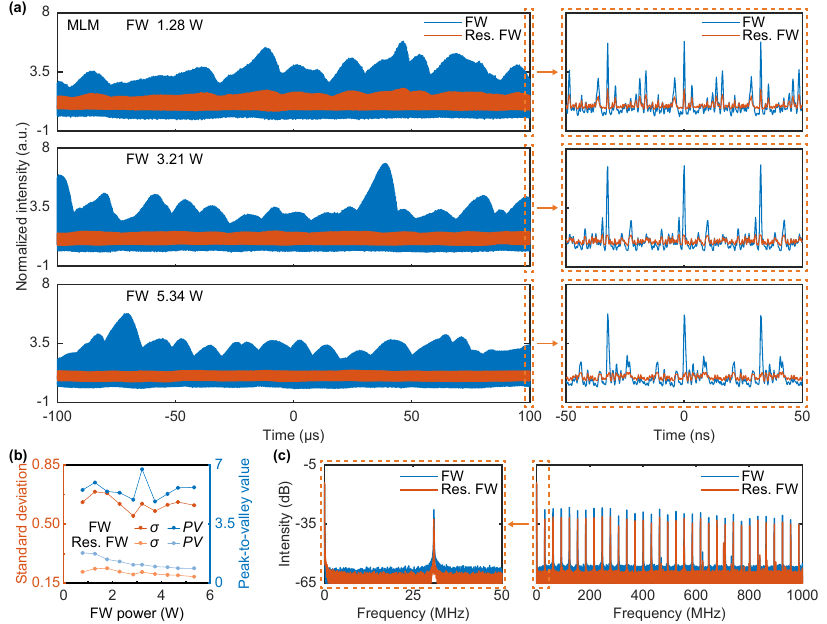}
\caption{\label{fig3} Characteristics of FW and Res. FW for MLM regime. (a) Normalized temporal profile of the FW and Res. FW in the range of 200 $\mu$s (left) and 100 ns (right) under different FW powers. (b) Standard deviations and peak-to-valley values of the FW and Res. FW varies with FW powers. (c) RF spectra of the FW and Res. FW. The inset of (c) shows a zoom-in view around the fundamental repetition rate.}
\end{figure}

When the pump current of seed laser is further increased to beyond 300 mA, the laser output transitions into the MLM regime. Figure~\ref{fig3}(a) presents normalized temporal profiles at three different FW powers, revealing that the temporal fluctuations in the residual FW progressively diminish with increasing FW power, demonstrating enhanced temporal beam self-cleaning effects. Furthermore, from the detailed inset of the temporal profile in the right panel, the SHG process exhibits the strongest absorption at the region where the FW intensity is highest. Figure~\ref{fig3}(b) shows the \(Std\) and \(PV\) values of both the FW and residual FW derived from normalized temporal profile at various FW powers. As illustrated, the \(Std\) and \(PV\) values of the residual FW decrease overall with increasing FW power, consistent with the trend observed in the normalized temporal profiles in Fig.~\ref{fig3}(a). At the maximum FW power, the \(Std\) decreases from 0.6122 for the FW to 0.189 for the residual FW, and the \(PV\) value drops from 5.6846 to 0.8847, representing approximately threefold and sixfold reductions, respectively. The strengthening of temporal stability at higher FW power levels is likely attributed to the increase in SHG efficiency corresponding to higher FW power. Figure~\ref{fig3}(c) displays the RF spectra under the maximum FW power of 5.34 W, where the residual FW exhibits significant reduction in beat frequency amplitude compared to the FW, providing evidence for the temporal stabilization mechanism achieved through SHG process.

\begin{figure}[b]
\includegraphics{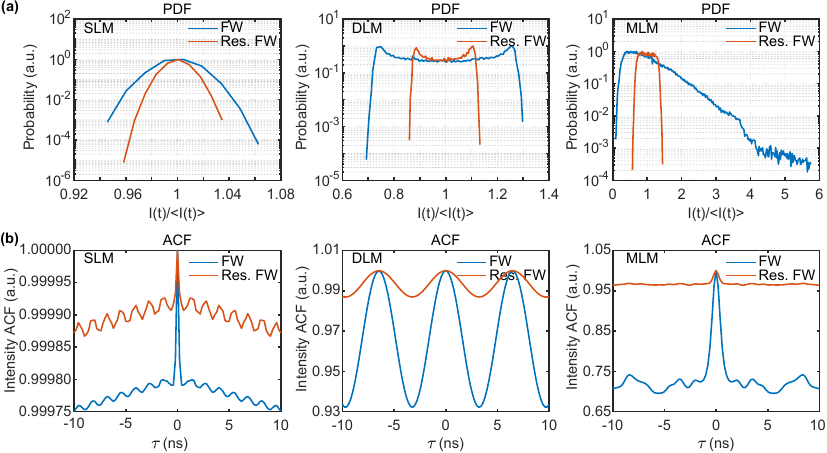}
\caption{\label{fig4} Impact of SHG process on the statistical characteristics of the laser source. (a) PDFs and (b) intensity ACFs of FW and Res. FW for SLM (left), DLM (middle), and MLM (right) regimes.}
\end{figure}

Furthermore, the influence of SHG process on the statistical properties of the laser source is investigated. Figure~\ref{fig4}(a) displays the PDFs for the FW and residual FW across three longitudinal mode regimes at their respective maximum FW power. As evident from the figure, the distribution range of normalized temporal intensity (defined as \(I(t)\)/\(<I(t)>\)) for the residual FW is narrower than that of the FW in all three longitudinal mode regimes. Statistically, for the SLM and DLM regimes, the normalized temporal intensity range at the probability of -20 dB decreased from $\sim$0.09 for FW to $\sim$0.05 for residual FW (a 44.4$\%$ relative change) and from $\sim$0.6 for FW to $\sim$0.27 (a 55$\%$ relative change) for residual FW respectively. While for the MLM regime, a dramatic decline from 3.38 to $\sim$0.81 occurred, corresponding to a significant 76$\%$ relative reduction. Figure~\ref{fig4}(b) presents the evolution of intensity ACF, $K(\tau) = \left \langle I(t) \cdot I(t+\tau)\right \rangle_t / \left \langle I(t)^2\right \rangle_t$, with delay time for FW and residual FW across all three longitudinal mode regimes at their respective maximum FW power. Evidently, the background level of the residual FW intensity ACF approaches 1 more closely than that of the FW across all three longitudinal mode regimes, demonstrating that the SHG process effectively strengthens the correlation of laser source. Specifically, the background level increases from $\thicksim$0.9998 to $\thicksim$0.9999 for the SLM regime and from $\thicksim$0.93 to $\thicksim$0.99 for the DLM regime when transitioning from the FW to the residual FW. For the MLM regime, the improvement is much more significant, rising from $\thicksim$0.72 to $\thicksim$0.96. These findings demonstrate that the SHG process brings the background level of the intensity ACF closer to unity, thereby strengthening the laser’s correlation properties and guiding it toward a coherent-state evolution.

The experimental findings discussed above confirmed the influence of SHG process on both temporal and statistical characteristics of the laser source. To further verify these observations and reveal the physical mechanisms of laser temporal stabilization, we performed numerical simulations to investigate SHG process on the characteristics of MLM fundamental wave sources. The coupled-wave equations for the SHG process can be expressed as follows \cite{flannigan2021}:

\begin{equation}\label{eq1}
       \begin{aligned}
              \left(\frac{\partial}{\partial z}+\frac{1}{v_{f}}\frac{\partial}{\partial t}\right)\overline{E}_{f}=id_\mathit{eff}\frac{\omega_{f}^{2}}{2c^{2}k_{f}}\overline{E}_{f}^{*}\overline{E}_{s}e^{-i\Delta kz},
       \end{aligned}
\end{equation}

\begin{equation}\label{eq2}
       \begin{aligned}
              \left(\frac{\partial}{\partial z}+\frac{1}{v_{s}}\frac{\partial}{\partial t}\right)\overline{E}_{s}=id_\mathit{eff}\frac{\omega_{s}^{2}}{4c^{2}k_{s}}\overline{E}_{f}^{2}e^{i\Delta kz}.
       \end{aligned}
\end{equation}

\noindent where Eqs. (\ref{eq1}) and (\ref{eq2}) govern the evolution of the fundamental wave and the second harmonic in the nonlinear crystal, respectively. The subscripts \(f\) and \(s\) in the equations denote the fundamental wave and the second harmonic components, respectively. \(E\) signifies the evolution of complex amplitude as a function of position \(z\) and time \(t\). $\omega$ denotes the angular frequency, \(k\) represents the wave propagation constant, and \(c\) is the speed of light in vacuum. \textit{v} represents the group velocity in the crystal, \(d_\mathit{eff}\) is the effective nonlinear coefficient of the nonlinear crystal, and $\Delta$\(k\) is the phase mismatch factor.

The fundamental wave refers to laser before undergoing SHG process, quantified by \(E_f\) at \(z=0\). In the simulations, \(E_f\) corresponds to the constructed fundamental wave for the DLM and MLM regimes. Simulation results for the MLM case are presented in the main text, whereas those for the DLM cases are provided in the Supplementary Document (refer to Supplementary Note S2 for details). Regarding the residual FW, it is obtained by taking the value of \(E_f\) at \(z=L\), where \(L\) is the length of the PPLN crystal. The parameter values used in the numerical simulation are listed in Table S1.

\begin{figure}
\includegraphics{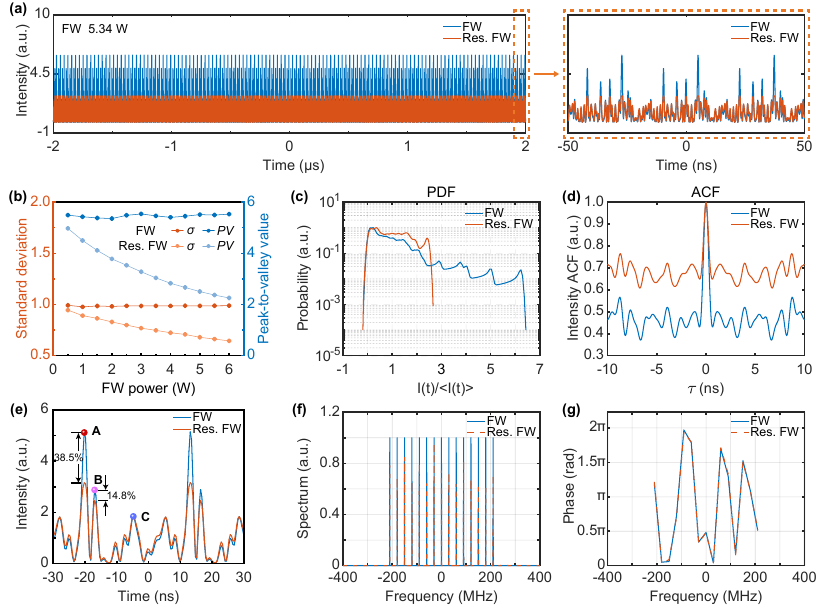}
\caption{\label{fig5} Simulation results regarding the influence of SHG process on the MLM laser source. (a) Normalized temporal profile of the FW and Res. FW in the range of 4 $\mu$s (left) and 100 ns (right) at the FW power of 5.34 W. (b) Standard deviations and peak-to-valley values of the FW and Res. FW varies with FW powers. (c) PDFs and (d) intensity ACFs of FW and Res. FW at 5.34 W FW power. (e) Detailed temporal profiles, (f) spectra, and (g) phase of the FW and the Res. FW.}
\end{figure}

Figure~\ref{fig5} displays the simulation results investigating the impact of SHG process on the MLM fundamental wave. Figure~\ref{fig5}(a) depicts the normalized temporal profiles of the FW and the residual FW in the MLM regime. Clearly, the temporal characteristics of the residual FW after SHG are significantly improved. Furthermore, the inset on the right panel demonstrates that the absorption strength increases with rising FW intensity during the SHG process, resulting in the stabilization of the temporal characteristics of the laser source. This observation further validates the earlier experimental results. Subsequently, we investigate the \(Std\) and \(PV\) value vary with FW power. During the temporal construction of the MLM regime, since the random phases of each longitudinal mode leads to substantial differences in the temporal profile across simulations. These differences cause the variations in the \(Std\) and \(PV\) value derived from each individual simulation, introducing strong randomness into the simulation results. To mitigate this problem, 300 independent simulations were performed for each FW power, and the mean value was adopted as the final result. Figure~\ref{fig5}(b) shows the \(Std\) and \(PV\) values across various FW power. The \(Std\) and \(PV\) value gradually decrease as the FW power increases, providing further evidence for the temporal stabilization effect induced by the SHG process. Figure~\ref{fig5}(c) depicts the PDFs of the FW and the residual FW at an FW power of 5.34 W. The figure reveals that the distribution of the normalized temporal intensity for the residual FW is narrower than that for the original FW, which is consistent with the experimental results. Figure~\ref{fig5}(d) displays the intensity ACFs of the FW and the residual FW at a FW power of 5.34 W, where the background level increases from $\thicksim$0.5 for the FW to $\thicksim$0.7 for the residual FW. It is evident that the background level of the residual FW is nearer to 1 than that of the FW, signifying that the residual FW is evolving towards coherent state.

To elucidate the physical mechanism underlying the temporal stabilization effect, Figs.~\ref{fig5}(e)-5(g) further presents a comparative analysis of the temporal details, spectra, and phase distributions of the FW and the residual FW. To avoid overlapping temporal pulses and difficulty in distinguishing them due to excessive longitudinal modes, the number of longitudinal modes in this simulation is set to 15 (fewer than the previous simulation setting) to present the temporal evolution details more clearly. Figure~\ref{fig5}(e) presents the normalized temporal comparison between the FW and the residual FW. As shown in the figure, the most significant weakening occurs at the highest peak of the fundamental wave (corresponding to point A), with an intensity decrease of about 38.5$\%$. The pulse with a lower peak (point B) decreases by about 14.8$\%$, and the pulse with an even lower peak (point C) remains almost unchanged. This indicates that pulses with higher peak power experience stronger absorption during the SHG process, exhibiting a clear ``high peak pulse suppression'' characteristic. To explain this phenomenon, the spectral and phase distributions of the FW and the residual FW are further provided, as shown in Figs.~\ref{fig5}(f) and \ref{fig5}(g). The results show that the amplitude of each longitudinal mode undergoes varying degrees of attenuation, but the phase of the laser before and after frequency doubling remains consistent. These two factors combined lead to the more pronounced suppression of the pulse with the higher initial peak during the SHG process.

Based on both experimental and numerical analyses, we demonstrate that the SHG process markedly enhances the temporal stability of the laser source while concurrently improving its correlation. This study demonstrates that the SHG process significantly suppresses intensity fluctuations and improves the temporal stability of the fundamental wave. These findings have important implications for the design of high-stability laser systems. Unlike conventional feedback-based methods that often sacrifice spectral purity \cite{tianNovelStructureRaman2023}, the SHG-based approach simultaneously strengths both temporal stability and correlation, making it particularly suitable for applications requiring narrow linewidth and low noise, such as coherent beam combining and precision metrology. Additionally, this approach faces significantly fewer limitations compared to material-based RSA effect implementations for temporal optimization.

\section{\label{sec:level1}Conclusions}
In summary, we investigated the influence of the SHG process on the laser source, examining the evolution of its temporal and statistical characteristics under SLM, DLM, and MLM regimes. A temporal comparison between the FW and the residual FW revealed that the SHG process enhances temporal stability of the laser source across three longitudinal mode regimes, with this stabilizing effect being most evident under MLM condition. Both experimental and numerical results demonstrate that the realization of temporal stabilization effect primarily originates from a pronounced suppression of high-peak-intensity pulses in the fundamental wave. Further numerical analysis reveals the underlying physical mechanism: the amplitudes of the individual longitudinal modes in the fundamental wave experience varying degrees of attenuation, while their phases remain invariant before and after SHG process. The interplay between these amplitude and phase characteristics ultimately results in a “high-peak-pulse compression” effect. Statistically, across all longitudinal mode regimes, the residual FW exhibits lower \(Std\) and \(PV\) values compared to the FW at arbitrary FW power, confirming the temporal stabilization achieved through SHG process. Particularly in the MLM regime, the \(Std\) and \(PV\) values of the residual FW decrease significantly as the FW power increases, as confirmed by simulations, which may be attributed to the enhanced SHG efficiency. At a maximum FW power of approximately 5 W, the \(Std\) and \(PV\) values of the residual FW are reduced by approximately threefold and sixfold compared with those of the FW. Analysis of the PDF and the intensity ACF shows that the SHG process improves the correlation properties of the light source, facilitating the evolution of the residual FW toward a coherent state. This study could advance comprehension of second harmonic generation and provide a novel method for enhancing temporal stability in laser source.

\begin{acknowledgments}
This work was supported by the National Key Research and Development Program of China (NKRDPC) (2022YFB3606000), the National Natural Science Foundation of China (NSFC) (62305391, 62275272), and the Innovation Reserch Foundation of NUDT (ZK 23-24).
\end{acknowledgments}

\section*{Author Contributions}
J.Y. and P.Z. conceived the idea of this study. S.C., L.D. and W.C. carried out the experiments. S.C. and J.Y. developed the numerical simulations. S.C., J.X., R.S. wrote the manuscript and analyzed the results. Z.J. supervised the project. The manuscript was written through contributions of all authors. All authors have given approval to the final version of the manuscript.

\bibliographystyle{apsrev4-2}

\bibliography{Reference}

\end{document}